\def\beq{\begin{equation}}
\def\eeq{\end{equation}}
\def\blg{\begin{align}}
\def\elg{\end{align}}
\def\beg{\begin{gather}}
\def\eeg{\end{gather}}
\def\bea{\begin{eqnarray}}
\def\eea{\end{eqnarray}}
\def\bed{\begin{displaymath}}
\def\eed{\end{displaymath}}
\def\bef{\begin{figure} \begin{center}}
\def\eef{\end{center} \end{figure}}

\def\1{\'{\i}}
\documentclass[preprint,pre,aps]{revtex4-1}
\usepackage[utf8]{inputenc}
\usepackage{graphics,graphicx,amssymb,amsmath,dcolumn,bm,url,amsbsy,pgf}

\begin{document}
\title{Quantifying coherence of chimera states in coupled chaotic systems}
\author{Carlos A. S. Batista $^1$ and Ricardo L. Viana $^2$ \footnote{Corresponding author. e-mail: viana@fisica.ufpr.br}}
\affiliation{1. Centro de Estudos do Mar, Universidade Federal do Paran\'a, Curitiba, Paran\'a, Brazil \\
2. Departamento de F\'isica, Universidade Federal do Paran\'a, Curitiba, Paran\'a, Brazil}

\date{\today}

\begin{abstract}
Chimera states in coupled oscillator systems present both spatially coherent and incoherent domains. The number and size of these domains depend on many factors like the system parameters and initial conditions. Systematic investigations of these dependences require a quantification of the degree of coherence present in a given snapshot spatial pattern. We propose the use of a local order parameter magnitude combined with the counting of the corresponding plateaus so as to provide such quantification. We use this technique in non-locally coupled lattices of chaotic logistic maps and chaotic R\"ossler systems to investigate the dependence of the degree of coherence on the coupling strength.
\end{abstract}

\maketitle

\section{Introduction}

Chimera states are statiotemporal patterns in arrays of identical coupled oscillators, for which incoherent and nonsynchronized domains coexist with coherent and synchronized ones \cite{abrams}. Even before its naming, chimera states were already observed in arrays of forced coupled Duffing oscillators \cite{umberger} and nonlocally coupled complex Ginzburg-Landau equations \cite{kuramoto}. However, nonlocal coupling is not the only way to achieve chimera states, since they can also occur due to spatially modulated delayed feedback coupling \cite{omelchenko}. 

The presence of chimera states has been identified in network models \cite{omelchenko,5} like the Kuramoto model of coupled oscillators \cite{6}, neuronal networks \cite{7} and lattices of coupled van der Pol-Duffing oscillators \cite{8}. Chimera states were also observed in experiments involving a liquid-crystal spatial light modulator, which controls the polarization properties of an optical wavefront, and which is an experimental realization of a nonlocally coupled map lattice \cite{aaron}. Other observations of chimera states are in populations of coupled chemical oscillators \cite{9}, coupled metronomes \cite{10,11}, and electronic oscillators \cite{12}.

The dynamics of chimera states has been investigated through a soluble model in Ref. \cite{abrams3}. For a one-dimensional chain of coupled oscillators, chimera states are typically characterized by a coherent domain next to an incoherent one \cite{abrams2}. The coherent domain, however, needs not to be synchronized, whereas the incoherent domain is always nonsynchronized. This fact presents a further difficulty in the quantitative characterization of chimera states, since there are so many different realizations of a coherent state that it is a nontrivial task to present a workable definition of the length of a coherent state or, what is equally difficult, to define the length of a incoherent state. We found that a suitable definition of length, when applied to chimera states, is provided by the so-called local order parameter introduced by Wolfrum {\it et al.} \cite{wolfrum,iryna}.

In this paper we use the local order parameter to introduce a suitable definition for the lengths of coherent and incoherent states in a non-locally coupled chaotic map lattice displaying chimera states. We use this definition to characterize quantitatively the degree of coherence, i.e. the relative size of the coherent domains with respect to the incoherent ones. This helps us to investigate the evolution of chimera states in a model system as its coupling parameters are varied. The methodology we present in this paper allows one to identify the coherent and incoherent groups of coupled oscillators in a given network, allowing the definition of an average quantity indicative of the length of the coherent plateaus. The method is thus able to: (i) be used to specify critical parameters defining the borders between the total coherent state, the total incoherent state and the chimera; (ii) estimate the critical parameter for the emergence of the chimera state in networks with infinite nodes, in the thermodynamic limit. 

This paper is organized as follows. Section II introduces the quantifiers to be considered in this work by investigation of a coupled chaotic logistic map lattice with finite range coupling. Section III considers a chain of coupled continuous-time R\"ossler systems. The last Section contains our conclusions. 

\section{Coupled chaotic map lattice}

One class of systems in which chimera states have been extensively studied is a one-dimensional chain of coupled chaotic maps with nonlocal coupling in the form \cite{iryna}
\begin{equation}
 \label{cml}
 x_{n+1}^{(i)} = f \left(x_n^{(i)}\right) + \frac{\varepsilon}{2P} \sum_{j=i-P}^{i+P} \left\lbrack f \left(x_n^{(j)}\right) - f \left(x_n^{(i)}\right) \right\rbrack,
\end{equation}
where $x_n^{(i)}$ is the state variable at discrete time $n$ and belonging to a chain of $N$ identical systems with periodic boundary conditions: $x_n^{(i\pm N)} = x_n^{(i)}$, with $i=1, 2, \ldots N$; $\varepsilon$ stands for the coupling strength, and $P$ denotes the number of neighbors coupled with a given map in both sides (finite-range coupling). Accordingly, $r=P/N$ is the coupling radius, which varies from $1/N$ (local or nearest-neighbor coupling) to $1/2$ (global or all-to-all coupling). 

We suppose that the local dynamics is governed by the logistic map $f(x) = ax(1-x)$, with $a=3.8$, for which the uncoupled maps exhibit chaotic behavior (this value will be held constant unless stated otherwise). It is known that, if the coupling is local (small coupling radius $r$) there is high-dimensional space-time chaos \cite{kaneko}. For non-local coupling, however, there have been observed many chimera states, with different wave numbers for the coherent domains, as the coupling parameters (the radius $r$ and the strength $\varepsilon$) are varied \cite{iryna}. 

\begin{figure}
\begin{center}
\includegraphics[width=0.6\textwidth,clip]{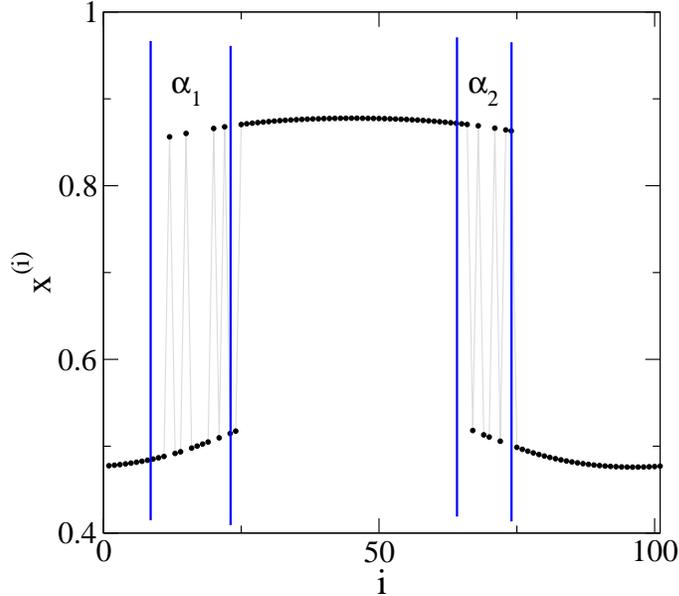}
\caption{\label{1} Snapshots of the spatial pattern for $N=100$ coupled chaotic logistic maps ($a=0.38$) with coupling radius $r = 0.32$ and coupling strength $\varepsilon = 0.32$. There have been used $10^5$ map iterations and random initial conditions.}
\end{center}
\end{figure}

A representative example is shown in Fig. \ref{1}, where we depict a snapshot of the spatial pattern for $N=100$ maps with coupling radius $r = 0.32$ and strength $\varepsilon = 0.32$. In this case there are two coherent states (or rather a single state with upper and lower branches) separated by two narrow layers of incoherent states, which we called $\alpha_1$ and $\alpha_2$. These incoherent layers become narrower as the coupling strength is increased, until they disappear at $\varepsilon = 0.43$. Similarly these layers become wider as $\varepsilon$ is decreased until all the chain has became incoherent as a whole \cite{iryna}

In order to quantify this transition from total coherence to total incoherence we use the local order parameter introduced in Ref. \cite{wolfrum}. Let ${\max}_j \{ x^{(j)} \}$ and ${\min}_j \{ x^{(j)} \}$ be, respectively, the maximum and minimum values of the state variable in a snapshot spatial pattern like Fig. \ref{1} (i.e. we take a fixed value of the discrete time $n$). We introduce a phase for the $j$th map from the following definition \cite{iryna}
\begin{equation}
 \label{psi}
 \sin\psi_j = \frac{2 x^{(j)} - {\max}_j \{ x^{(j)} \} - {\min}_j \{ x^{(j)} \}}{{\max}_j \{ x^{(j)} \} - {\min}_j \{ x^{(j)}\} }, \qquad (j=1, 2, \ldots N)
\end{equation}
such that a spatial half-cycle is mapped onto the phase interval $[-\pi/2,\pi/2]$. The local order parameter magnitude is defined as 
\begin{equation}
 \label{order}
 R_k = \lim_{N\rightarrow\infty} \frac{1}{2\delta(N)} \left\vert \sum_{j\in C} e^{i\psi_j} \right\vert, \qquad (k=1, 2, \ldots N)
\end{equation}
where the summation is restricted to the interval of $j$-values such that
\begin{equation}
 \label{condition}
 C: \left\vert \frac{j}{N} - \frac{i}{N} \right\vert \le \delta(N),
\end{equation}
where $\delta(N)\rightarrow 0$ for $N\rightarrow\infty$. 

\begin{figure}
\begin{center}
\includegraphics[width=0.7\textwidth,clip]{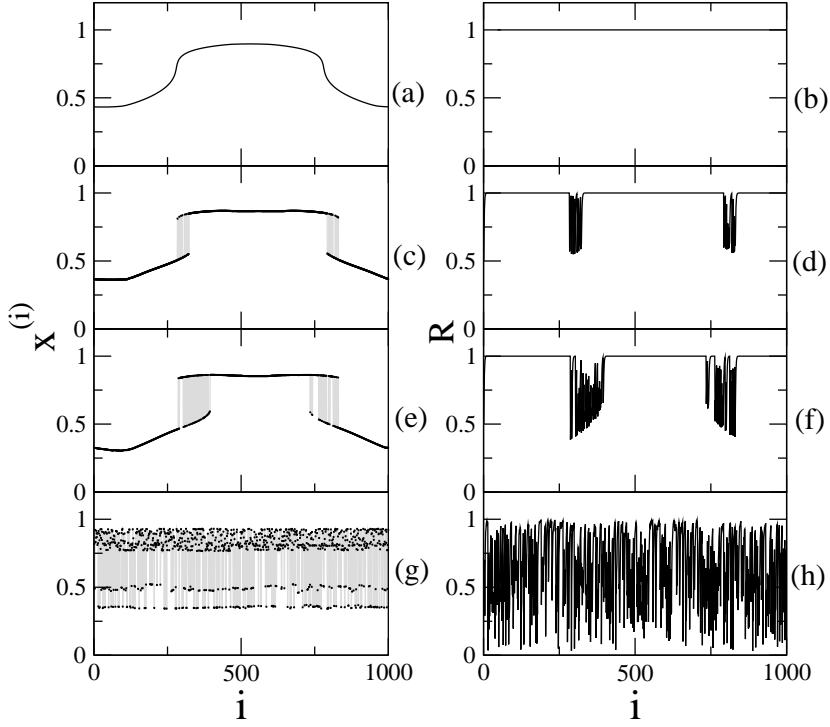}
\caption{\label{2} Snapshots of the spatial pattern for $N=1000$ coupled chaotic logistic maps ($a=0.38$) with coupling radius $r = 0.32$ and coupling strengths $\varepsilon =$ (a) $0.43$; (c) $0.35$; (e) $0.30$ and (g) $0.00$. (b), (d), (f), and (h) are the local order parameter magnitudes corresponding to (a), (c), (e), and (g), respectively.}
\end{center}
\end{figure}

It turns out that $R_i \approx 1.0$ for sites belonging to coherent domains, and takes on lesser values for sites within incoherent domains in a chimera state. In Fig. \ref{2} we illustrate the transition from coherent to incoherent states in a lattice of $N=1000$ maps with coupling radius $r = 0.32$ and coupling strength $\varepsilon$ varying from zero to $0.43$. In the former case the snapshot at $\varepsilon = 0.43$ shows that the whole lattice is a coherent domain [Fig. \ref{2}(a)], and the corresponding local order parameter magnitude is accordingly close to the unity for all lattice sites [Fig. \ref{2}(b)]. Decreasing the coupling strength we observe two narrow layers of incoherent behavior [Fig. \ref{2}(c)], for which the values of $R_i$ decrease with respect to $1.0$ [Fig. \ref{2}(d)]. The widths (and also the positions) of the incoherent layers increase with decreasing $\varepsilon$ [Fig. \ref{2}(e)], what can be also observed in the local order parameter magnitude values [Fig. \ref{2}(f)]. Finally, for uncoupled maps the whole lattice becomes incoherent [Fig. \ref{2}(g)] and the values of $R_i$ are likewise fluctuating randomly between zero and unity [Fig. \ref{2}(h)]. 

\begin{figure}
\begin{center}
\includegraphics[width=0.5\textwidth,clip]{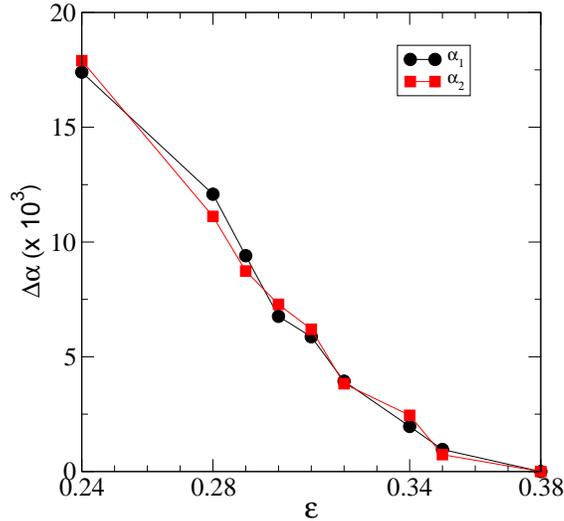}
\caption{\label{3} (color online) Widths of the incoherent regions as a function of the coupling strength for $N=8 \times 10^5$ coupled chaotic logistic maps ($a=0.38$) with coupling radius $r = 0.33$.}
\end{center}
\end{figure}

From the sizes of the local order parameter plateaus it is straightforward to evaluate the widths of both coherent and incoherent domains, whenever they coexist in a chimera state. We applied this procedure to a large number of snapshots obtained with the same coupling radius of $r = 0.33$, from $\varepsilon = 0.24$ to $0.38$, and computed the length of the incoherent states $\alpha_1$ and $\alpha_2$, denoted $\Delta\alpha$, the results being shown in Fig. \ref{3} as a function of $\varepsilon$ for $\alpha_1$ and $\alpha_2$. Both regions have roughly the same width, which decreases with growing $\varepsilon$, as we have already antecipated from Fig. \ref{2}: when $\varepsilon > 0.38$ the widths of both $\alpha_1$ and $\alpha_2$ go to zero and the whole lattice becomes coherent. On the other hand, as $\varepsilon = 0.24$ the sum of the widths is comparatively large and the lattice becomes mostly incoherent. This method, however, may not be accurate enough to give a critical value for the onset of a chimera state. Accordingly we need a more appropriate diagnostic of coherence for this system.

A general feature of Fig. \ref{2} is that, within a coherent domain, the local order parameter magnitude has a plateau at a value very close to $1.0$. We introduce a degree of coherence $p$ for a chimera state by computing the relative mean plateau size for a given snapshot pattern \cite{sandro}. In the general case we can have more than one coherence plateau, so let $N_i$ be the length of the $i$th plateau, and $N_p$ the total number of plateaus. The mean plateau size is thus ${\tilde N} = (1/N_p) \sum_{i=1}^{N_p} N_i$. The coherence degree is the ratio between the mean plateau size and the total lattice size, or $p = {\tilde N}/N$. If the whole lattice is totally coherent, like in Fig. \ref{2}(a), we have just one plateau and ${\tilde N}=N$, hence $p=1$. On the contrary, if the lattice is completely incoherent, like in Fig. \ref{2}(g), there are as many plateaus as maps, or $N_p \approx N$, which yields ${\tilde N} \approx 1$ and $p \approx 1/N \rightarrow 0$ as $N\rightarrow\infty$. 

\begin{figure}
\begin{center}
\includegraphics[width=1.0\textwidth,clip]{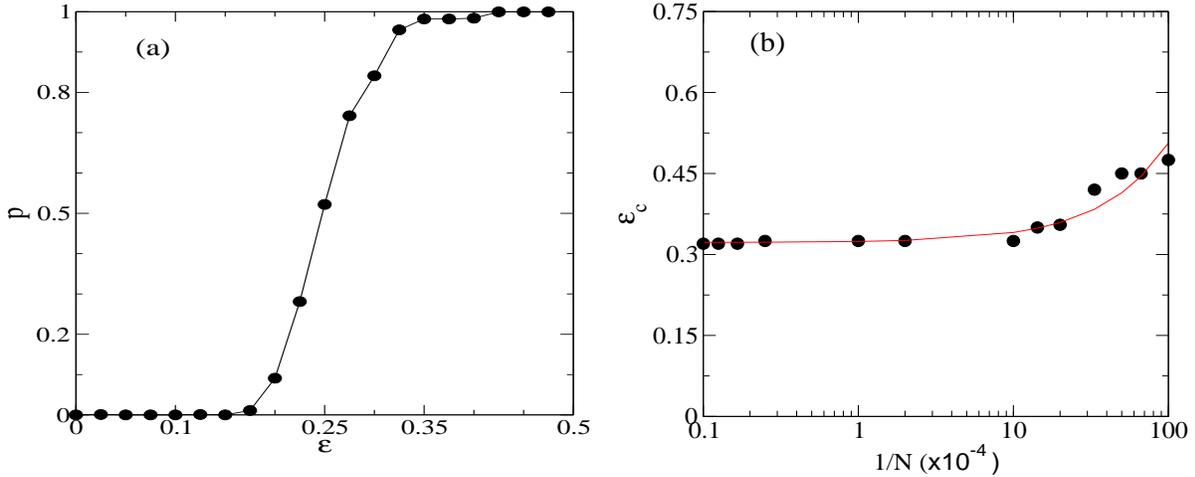}
\caption{\label{4} (a) Degree of coherence as a function of the coupling strength, and (b) critical value of the coupling strength for the formation of chimera state as a function of the inverse lattice size, for $N=1000$ coupled chaotic logistic maps ($a=0.38$) with coupling radius $r = 0.33$.}
\end{center}
\end{figure}

In Fig. \ref{4}(a) we plot the coherence degree as a function of the coupling strength $\varepsilon$, showing that the chimera state (existence of partially incoherent domains) is observed, starting from a completely coherent domain, as $\varepsilon$ is decreased from the critical value $\varepsilon_c \approx 0.32$. In Fig. \ref{4}(b) we plot the value of $\varepsilon_c$ as a function of the inverse lattice size, showing a slight decrease as the number of coupled maps increases. Extrapolating to the thermodynamical limit of $N\rightarrow\infty$ we find that ${(\varepsilon_c)}_{\infty} \approx 0.322$. 

\section{Chains of coupled R\"ossler systems}

Chimeras can also be readily found in lattices of non-locally coupled continuous-time flow, like the R\"ossler system. Using a finite-size coupling analogous to (\ref{cml}) we have also studied the following spatially extended system
\begin{align}
 \label{r1}
 \frac{dx_i}{dt} & = - y_i - z_i + \frac{\varepsilon}{2P} \sum_{j=i-P}^{i+P} (x_j - x_i), \\
 \label{r2}
 \frac{dy_i}{dt} & = x_i + a y_i + \frac{\varepsilon}{2P} \sum_{j=i-P}^{i+P} (y_j - y_i), \\
 \label{r3}
 \frac{dz_i}{dt} & = b + z_i (x_i - c) + \frac{\varepsilon}{2P} \sum_{j=i-P}^{i+P} (z_j - z_i), 
\end{align}
where we choose $a = 0.42$, $b = 2.0$ and $c=4.0$ so as to give chaotic attractors for the uncoupled systems. 

\begin{figure}
\begin{center}
\includegraphics[width=0.8\textwidth,clip]{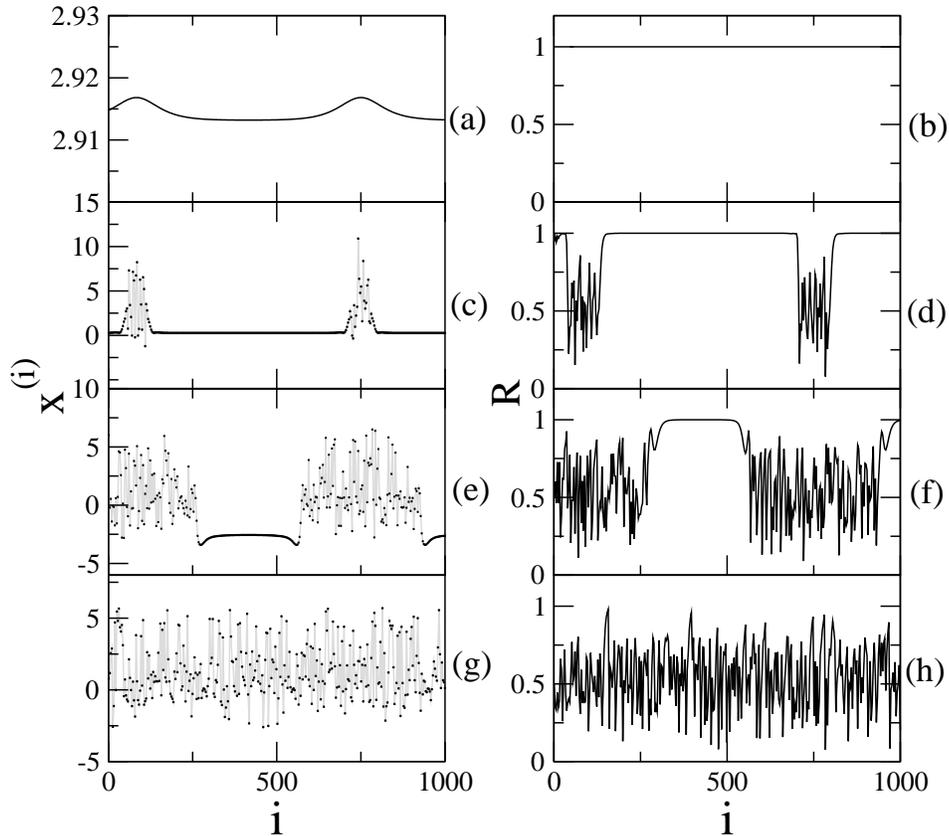}
\caption{\label{8} Snapshots of the $x$-variable for a lattice of $N=1000$ R\"ossler systems with coupling radius $r=0.3$ and (a) $\varepsilon = 0.39$, (c) $\varepsilon=0.37$, (e) $\varepsilon=0.35$, and  (g) $\varepsilon=0.20$. The corresponding profiles for the local order parameter magnitude are depicted in (b), (d), (f), and (g), respectively. The snapshots were taken at fixed time $23,000$ using periodic boundary conditions and sinusoidal initial conditions.}
\end{center}
\end{figure}

According to the snapshots depicted in Fig. \ref{8}, obtained for lattices of $N = 1000$ R\"ossler systems with coupling radius $r=0.3$, we can see an evolution of a coherent (almost completely synchronized) state [Fig. \ref{8}(a), obtained for $\varepsilon = 0.39$] to an incoherent (almost completely non-synchronized) state [Fig. \ref{8}(g), obtained for a smaller coupling strength of $\varepsilon = 0.20$], showing chimeras for intermediate values of $\varepsilon$ [Figs. \ref{8}(c) and (e), obtained for $\varepsilon=0.37$ and $0.35$, respectively]. Just like in the previous example of coupled chaotic map lattices, the local order parameter magnitude is able to evidence the coherent and incoherent parts of the snapshot profile [Figs. \ref{8}(b), (d), (f) and (g)]. 

\begin{figure}
\begin{center}
\includegraphics[width=1.0\textwidth,clip]{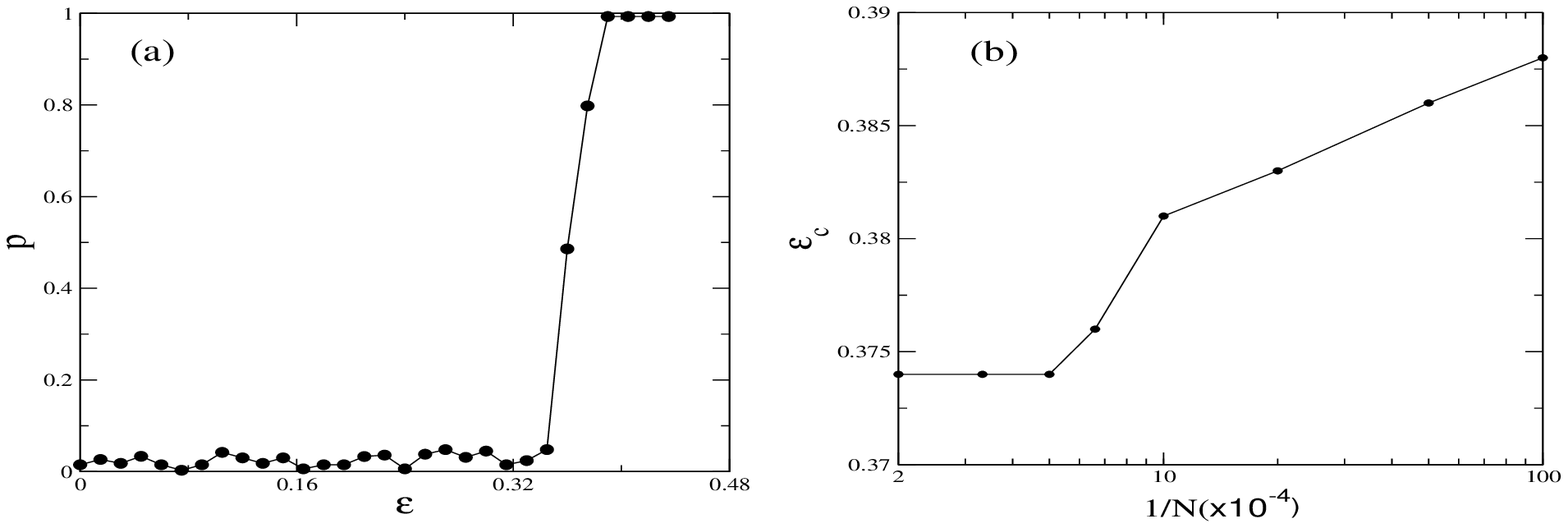}
\caption{\label{7} (a) Degree of coherence as a function of the coupling strength for a lattice of $N=1000$ R\"ossler systems with coupling radius $r=0.3$. (b) Critical value of the coupling strength as a function of the inverse lattice size a lattice of $N=1000$ R\"ossler systems with coupling radius $r=0.3$.}
\end{center}
\end{figure}

The sequence snapshots displayed in Figure \ref{8} suggests that, for this particular set of values, there is a transition from a coherent to an uncoherent state as the coupling strength is decreased past a critical value $\varepsilon_c$, as revealed by Fig. \ref{7}(a), where we show the evolution of the degree of coherence $p$ with $\varepsilon$, indicating a transition for $\varepsilon_c \approx 0.38$. As in the previous example, we also observed that the value of $\varepsilon_c$ decreases as the lattice size is increased [Fig. \ref{7}(b)], and tends to $\sim 0.374$ in the limit $N \rightarrow \infty$.

\section{Conclusions}

In conclusion, we have considered in this paper a quantitative characterization of chimera size and evolution in lattices of nonlocally coupled dynamical systems. We have considered, as illustrative examples, dynamical systems in both discrete (chaotic logistic maps) and continuous time (R\"ossler systems), and we believe that our results will hold for any spatially extended dynamical systems, even for partial differential equations. We have investigated coupled systems that presents both spatially coherent and incoherent domains (chimera states) such that, for selected sets of parameter values, there is a dependence of the sizes of the incoherent states with the coupling strength. If this parameter is varied within a certain range, the incoherent and coherent domain lengths are changed, maintaining fixed values of the other parameters of the system. 

By using a degree of coherence definition based on plateaus of the local order parameter magnitude, the numerical results showed three behavior regimes. First, if the coupling strength is relatively small, coherent domains are absent. Second, a transition between coherent and incoherent domains appears due to the variation of coupling strength. Third, if the coupling is relatively large in the system, the coherent domains are predominant. We found that in the abovementioned transition, there exists a weak dependence on the size of network. This was confirmed through the computation of the critical value for the coupling strength. Nevertheless, the technique we propose to quantify the sizes of incoherent and coherent domains is general enough to be used in any similar system, both in discrete and continuous space and time, provided a suitable definition of a local order parameter magnitude is given. 

\section*{Acknowledgments}

This work has been partially supported by the Brazilian Government Agency CNPq.

\end{document}